\documentclass[conference,12pt]{IEEEtran}

\ifCLASSINFOpdf
\else
\fi

\usepackage[square, comma, sort&compress,numbers]{natbib}
\usepackage{flushend}
\usepackage{caption}
\usepackage{amsmath}

\usepackage{fancyhdr}
\usepackage{setspace}
\fancypagestyle{firststyle}
{
   \fancyhf{}
   \fancyfoot[C]{\footnotesize Technical Report, 2014.}
}

\onecolumn
\begin{document}
\bstctlcite{IEEEexample:BSTcontrol}

%\onecolumn
% paper title
% can use linebreaks \\ within to get better formatting as desired
\title{Power Management Techniques for Data Centers: A Survey}

\author{Sparsh~Mittal\\% <-this % stops a space
Future Technologies Group\\ Oak Ridge National Laboratory\\ Oak Ridge, Tennessee, USA 37830\\
mittals@ornl.gov

}% <-this % stops a space

% use for special paper notices
%\IEEEspecialpapernotice{(Invited Paper)}

% make the title area
\maketitle

\begin{abstract}
With growing use of internet and exponential growth in amount of data to be stored and processed (known as ``big data''), the size of data centers has greatly increased. This, however, has resulted in significant increase in the power consumption of the data centers. For this reason, managing power consumption of data centers has become essential. In this paper, we highlight the need of achieving energy efficiency in data centers and survey several recent architectural techniques designed for power management of data centers. We also present a classification of these techniques based on their characteristics. This paper aims to provide insights into the techniques for improving energy efficiency of data centers and encourage the designers to invent novel solutions for managing the large power dissipation of data centers.
\end{abstract}

% IEEEtran.cls defaults to using nonbold math in the Abstract.
% This preserves the distinction between vectors and scalars. However,
% if the journal you are submitting to favors bold math in the abstract,
% then you can use LaTeX's standard command \boldmath at the very start
% of the abstract to achieve this. Many IEEE journals frown on math
% in the abstract anyway.

% Note that keywords are not normally used for peerreview papers.
\begin{IEEEkeywords}
Data Centers, Power Management, Low-power Design, Energy Efficiency, Green Computing, DVFS, Server Consolidation

\end{IEEEkeywords}

\IEEEpeerreviewmaketitle
\thispagestyle{firststyle}

\section{Introduction}
As we move to the exascale era, the requirements of data storage have grown exponentially and hence, the power consumption of data-centers has also increased \cite{baliga2011green,kant2009data}. It has been reported that in year 2006 alone, the datacenters and servers in U.S. consumed 61 billion kilowatt hours of electricity, which is 1.5\% of all US electricity consumption and has a monetary cost of \$4.5 billion \cite{brown2008report}. Many data centers which support cloud services employ tens of thousands of servers which draw tens of Mega-Watts of power at peak \cite{greenberg2008cost}. It has been estimated that such data centers draw power worth \$9.3 million per year \cite{greenberg2008cost}. Moreover, the worldwide expenditure on enterprise power supply and cooling has been estimated to be more than \$30 billion \cite{patel2006enterprise}. Hence, achieving energy efficiency in data centers has become an issue of paramount importance. 

Recently, several techniques have been proposed for managing power consumption of data centers (e.g. \cite{bergamaschi2012data,cabusao2010data,chang2012totally,chen2012fine,chetsa2012beyond}). Due to large power consumption levels of data centers, use of these techniques has become essential to maintain both energy efficiency and cost efficiency. In this paper, we highlight the need of achieving energy efficiency in data centers and survey several techniques which have been proposed for ensuring green operation of data centers. Since it is not possible to include detailed discussion of a large number of techniques, we take the following approach to limit the scope of this paper. We focus on architectural-level and operation-level techniques for improving energy efficiency and not on circuit-level (device-level) techniques. Moreover, we only discuss the key design principles of different techniques and do not discuss the quantitative results, since different techniques have been tested using different evaluation platforms. We only include those techniques which have been evaluated for improving energy efficiency, and do not include the techniques aimed at improving performance, although they are also expected to improve energy efficiency. 

The rest of the paper is organized as follows. In the next section, we present the factors which lead to increased power consumption. We also discuss their impact along with the need of managing data center power consumption. We then present a classification of power management techniques based on their characteristics, and then discuss some of these power management techniques in detail. Finally, we provide a discussion of future research directions and concluding remarks.

\section{Background}
\subsection{Reasons behind Increase in Power Consumption}

Due to recent growth in use of internet, the demand placed on data-centers has increased. Modern data centers typically contain up to tens of thousands of servers and provide 24$\times$7 services to hundreds or thousands of users. As an example, YouTube serves as much as 100 million videos a day, while Facebook has nearly 400 million active users and 3 billion photos uploaded every month \cite{hepburn2011facebook}. These images, videos and text data are processed using computation-intensive software programs; and stored and accessed from data centers \cite{mittal2013bioinqa,mittal2011versatile}. Further, in recent years, use of high-performance computing techniques has also increased, which require trading off energy consumption for obtaining increased performance. For this reason, data centers have grown in sizes; which has led to increase in their power consumption.

\subsection{Power Consumption Levels}

Recent years have witnessed a huge increase in the power consumption of servers, data centers and supercomputers \cite{mittal2012survey,zhang2012green,MitZha13_Cashier} . The total energy consumption of data centers as a percentage of total US energy consumption has doubled between 2000 and 2007 \cite{kant2009data}. As an example, peak power consumption of the most powerful supercomputers in the TOP500 list of supercomputers ranges in tens of megawatts \cite{feng2007green500,top500supercomputer}. This amount of power is sufficient to fulfill the demands of a city of 40,000 residents. Research has also shown that the buildings with data centers can be as much as 40 times more energy intensive than the conventional office buildings \cite{greenberg2006best}.

The high levels of power consumption demand a costly cooling infrastructure \cite{mittal2014survey}. For example, a 30,000 square feet data center with 10 megawatts power consumption consumes on the order of \$5 million for cooling in a year \cite{patel2003smart}. Moreover, for every watt of power consumed in the computing equipment, an additional 0.5 to 1W of power is required to operate the cooling system itself \cite{patel2003smart}, which further adds to the cost.

\subsection{Need of Power Management}

It has been shown that most of the time, the modern servers operate between 10\% to 50\% of maximum possible utilization \cite{barroso2009datacenter,ranganathan2006ensemble}. Further, at these utilization levels, the server energy efficiency also becomes very low \cite{barroso2007case}. Thus, despite the fact that the average utilization remains low; there exist frequent, brief bursts of activity, and to meet the requirements of service-level-agreements (SLAs), operators are forced to allocate high amount of resources, which leads to poor energy efficiency \cite{MitZha12_EnCache}.

Power management is also important from an economic point of view, since effective power management also improves operational efficiencies and increases compaction. Electricity costs for powering servers forms a major cost of operation in data centers and it has been estimated that in near future, energy costs may contribute even more than the cost of IT (\cite{barroso2003web,bianchini2004power}). Further, a high ratio of cooling power to computing power restricts the compaction and consolidation possible in data centers, which results in increased operation costs. For example, the high power density poses significant challenges in routing the large amounts of power needed per rack. Currently the power delivery in typical data centers is near 60 Amps per rack and it is expected to reach the limit of power delivery, which will severely affect the operation of servers \cite{ranganathan2006ensemble}.

The large power consumption and high concentration of nodes in data centers leads to increased node failures. It has been observed that a 15 degree Celsius rise increases the failure rates in hard-disks by a factor of two \cite{anderson2003more}. Hence, maintaining the computer systems at proper temperature is important for ensuring maximum reliability, longevity, and large return on investment. 
Finally, large power consumption also has adverse environmental impact, e.g. large carbon emission  \cite{daim2009data,energystar,feng2003making,mazzucco2012empirical,sego2012implementing}. For these reasons, the design of green solutions for modern data centers has become a topic of paramount importance. Hence, in this paper, we survey techniques for managing power consumption in data centers.

 \section{Overview of Power Management Techniques}

In recent years, researchers have proposed several techniques for managing power consumption in data centers. While it is very difficult to draw sharp boundaries of classification; for the purpose of study, we broadly classify the techniques in the following four types.
\begin{enumerate}
\item DVFS (dynamic voltage/frequency scaling) based techniques \cite{bohrer2002case,deng2012multiscale,freeh2007analyzing,gandhi2009optimal,horvath2007dynamic,hsu2005power,rao2010minimizing,sharma2003power,shi2011unified,xu2005energy}.

\item Techniques which transition the server/node to low-power or turned-off state; or use server-consolidation based approach to allocate only required amount of server resources  \cite{anagnostopoulou2009energy,bohrer2001energy,chase2001managing,chen2011spatially,da2010multi,ghosh2012knightshift,liu2009greencloud,pinheiro2003dynamic,rajamani2003evaluating,ranganathan2006ensemble,verma2009server,zheng2009justrunit,leverich2009power,xu2005energy}.

\item  Workload management or task scheduling based techniques \cite{banerjee2011integrating,bradley2003workload,le2009cost,moore2005making,nathuji2007exploiting,wang2009towards,zapater2012leveraging,tang2006thermal}.

\item Thermal-aware or thermal-management techniques which take into account the thermal properties \cite{banerjee2011integrating,bash2006dynamic,huang2011tapo,kudithipudi2013thermal,mirtar2012adaptation,moore2005making,moore2004going,pakbaznia2010temperature,ramos2008c,wang2009towards}. Further, some techniques address the issues related to cooling in data centers \cite{kim2012free,patel2003smart,moore2005making}.
\end{enumerate}

Based on other characteristics/parameters, the techniques can be further classified. While most techniques aim to reduce average power (energy), a few techniques aim to reduce peak power consumption \cite{felter2005performance}, or limit power (called power capping) \cite{choi2008profiling,lefurgy2007server}. A few techniques aim to save energy while keeping the performance degradation bounded or maintaining QoS (quality-of-service) guarantee \cite{hsu2005power,kephart2007coordinating,kusic2009power,liu2009greencloud,rusu2006energy,sharma2003power,sudan2012optimizing,wang2009co,xu2005energy}.

Several techniques use analytical approach and offer control-theoretical algorithms with provable guarantees \cite{chen2005managing,horvath2007dynamic,horvath2008multi,lefurgy2007server,parolini2010cyber,urgaonkar2010dynamic,wang2008cluster,wang2009co}, while most other techniques use system approach and focus only on implementation.
Some researchers propose use of renewable energy sources \cite{goiri2011greenslot,nguyen2013powering}. Some researchers aim to reduce disk energy \cite{gurumurthi2003drpm,narayanan2008write,pinheiro2006exploiting}, while others focus on saving main memory energy \cite{chatterjee2012leveraging,deng2012multiscale,deng2011memscale,yoon2012boom,liu2012raidr}  in data centers.

\section{Power Management Techniques}

\subsection{DVFS Based Techniques}

Dynamic voltage and frequency scaling (DVFS) is a widely-used power-management technique where the clock frequency of a processor is dynamically adjusted to allow a corresponding reduction in the supply voltage to achieve power saving \cite{mittal2013survey}. DVFS is especially useful for memory-bound workloads. The power consumption of a CMOS (Complementary metal oxide semiconductor) circuit is given by
\begin{equation}
P = P_{static} + CFV^2 
\end{equation}

Here $C$ shows the capacitance of the transistor gates, $F$ shows the operating frequency and $V$ shows the supply voltage. The frequency at which circuit is clocked determines the voltage required for stable operation; and hence, by intelligently reducing the frequency, the supply voltage can also be reduced; which leads to significant power savings due to $V^2$ relationship shown above. The limitation of DVFS, however, is that a reduction in frequency also reduces the performance of the circuit and hence, DVFS may adversely affect the performance of the processor. For this reason, DVFS needs to be intelligently applied, to maintain high performance. In data centers, DVFS technique is applied to manage the power consumption of multicore processors, DRAM memories and other components.

Sharma et al. \cite{sharma2003power} propose adaptive algorithms for dynamic voltage scaling in QoS-enabled web servers. Their algorithms aim to minimize energy consumption subject to service level agreements (SLAs). The algorithms are implemented inside the Linux kernel. Their algorithms minimize the energy consumption using a feedback loop which regulates the frequency and voltage levels to keep the instantaneous utilization bounded.

Hsu and Feng \cite{hsu2005power} propose an algorithm for the dynamically varying voltage and frequency of the processor to realize energy saving while keeping the performance loss bounded. The algorithm takes decisions at the end of fixed time period. Their algorithm uses an estimation model to relate the intensity level of off-chip accesses to total execution time. Using this, the algorithm computes the lowest CPU frequency which keeps the performance loss bounded, while saving largest possible amount of energy. For multiprocessor environment, the same algorithm is repeated for each processor to set the frequency of the processor individually.

Xu et al. \cite{xu2005energy} propose a technique for saving energy in embedded clusters. Their technique adjusts the number of active nodes based on the system load. In their technique, each node in the cluster performs dynamic voltage scaling independently and runs at the lowest frequency at which it can keep up with the request arrival rate. To prevent the system from reacting to short-term changes in workload, at a time, their technique changes only one node from active to inactive and vice-versa in each interval.

Horvath et al. \cite{horvath2007dynamic} present methods to dynamically adjust the server voltages to minimize the total system power consumption, while also meeting end-to-end delay constraints in a multi-tier web service environment. In their architecture, server machines use DVS-capable processors. They propose coordinated distributed voltage scaling policy, where decisions on frequency adjustments are made on each locally while minimizing overall power consumption. Compared to other methods such as server on/off, use of dynamic voltage scaling incurs less overhead which facilitates aggressive energy saving. 

Deng et al. \cite{deng2011memscale} use DVFS mechanism to save memory energy. Their technique lowers the frequency of DRAM devices, memory channels and memory controllers at the time of low memory activity. This reduces the memory power consumption. They have also extended their technique for coordinating DVFS across multiple memory controllers, memory devices and channels to reduce the overall system power consumption \cite{deng2012multiscale}.

\subsection{Server Consolidation and Power State Transitioning Based Techniques}

As discussed before, modern servers typically operate at low utilization levels. Moreover, to cater to the peak demand and service-level-agreements and ensure reliability, high amount of server resources needs to be allocated which leads to poor energy efficiency \cite{barroso2007case}. To address this challenge, many approaches have been proposed. Server consolidation is one such approach aimed to ensure efficient usage of server resources by reducing the total number of servers required by a data center, while still delivering same throughput. In this approach, the existing applications are consolidated onto fewer servers, such that unused servers can be transitioned into low-power (or turned-off) state and the used servers can be operated at full utilization levels. Another approach is transitioning the server resources into low-power mode during periods of low activity. These approaches have been widely used to improve energy efficiency of data centers.

Chun et al. \cite{chun2010energy} propose a hybrid data center design which uses heterogeneous platforms to save power. Under low utilization levels, their technique transfers the running tasks from a high-power, high performance system to a low-power, low-performance system and turns off the higher power servers. Thus, server virtualization along with task migration reduces the number of active servers by consolidating many servers with low-average utilization to a few servers operating at high utilization. The limitation of sever consolidation, however, is that it may lead to slow response times and high transition costs.

For ensuring energy-efficient operation of data centers, Chase et al. \cite{chase2001managing} discuss a system resource management based approach. Their method controls server allocation and routing of requests to selected servers using a dynamically reconfigurable switch; and thus enables achieving a trade-off between service quality and cost. For optimizing energy consumption, incoming request traffic is continuously monitored and only desired amount of server resources are allocated such that the service level agreements can be fulfilled. Since internet sites observe highly varying usage patterns (e.g. as much as 11:1 peak-to-trough ratio of usage), their method enables intelligent adaptive resource provisioning. 

Ranganathan et al. \cite{ranganathan2006ensemble} propose a technique for managing server power at the ensemble (i.e. collection of systems) level instead of individual server level. Their design observes the resource-usage trends across multiple systems. By taking advantage of inter-server variations, their technique allows active server to steal power from the inactive servers. Their technique facilitates reductions in the requirements for power delivery, power consumption, and cooling in the data centers.

To leverage the variation in workload for saving energy, Anagnostopoulou et al. \cite{anagnostopoulou2009energy} propose a ``barely-alive'' server design. Their design transitions the servers to a barely-alive power state, where the server can be still accessed, even if many of its other components are turned off. Their design uses a small embedded processor to only keep the memory of idle servers active so that in-memory application code/data remain unaffected, and the free memory space can be used for cooperative application data caching.

To reduce the energy consumption of servers during periods of reduced load, Rusu et al. \cite{rusu2006energy} discuss a cluster-wide QoS-aware technique which uses dynamic reconfiguration based approach. For a given workload, their algorithm dynamically decides the servers which need to remain turned on/off to minimize global power consumption. To minimize the time penalty of server on/off, the algorithm accounts for the booting time of the server and turns on a server before it is actually needed. The authors also use dynamic voltage scaling (DVS) to conduct QoS-aware power management. The authors show the use of their technique in the context of a web server.

Ghosh et al. \cite{ghosh2012knightshift} propose a technique to save energy in data centers by using out-of-band management processors which are typically used for remotely managing a server, to satisfy the I/O requests from a remote server. By transferring the load from the primary server to the management processor, their technique allows the primary server to stay in low-power state for longer time, which improves the energy efficiency.

To utilize the opportunity of energy saving at multiple levels, Da Costa et al. \cite{da2010multi} present an integrated framework, called GREEN-NET. GREEN-NET provides a multi-faceted approach for saving energy in clouds and grids. Their approach has three levels. At first level, it increases the awareness of users of their energy consumption. At second level, it involves the users in decisions to trade-off performance for saving energy. Finally, at third level, it conducts adaptive management of grids by techniques such as server turn-off to save energy.

L. Liu et al. \cite{liu2009greencloud} use virtual-machine migration approach where a virtual machine (VM) is transferred across physical computers to enable server consolidation and allow more computers to be turned-off. Their technique dynamically decides the time to trigger VM migration, and the alternative physical machines to achieve optimal VM placement. Their technique enables live migration of VMs, such that the users can hardly notice that their applications are being or have been migrated.

Leverich et al. \cite{leverich2009power} use per-core power-gating (PCPG) approach to manage power consumption of multicore processors. Their technique works by selectively turning on/off the power supply to individual cores of a multicore processor based on the utilization and quality-of-service requirements. The authors have also shown that by syngersitically combining their technique with DVFS technique, additional power savings can be obtained.	 

\subsection{Workload Scheduling Based Techniques}

Modern data centers typically have a large number of servers and hence, the decision about placement of workloads on specific servers significantly affects the heat-dissipation and power-consumption. A poor placement may greatly increase the temperature of the building which will further increase the heat-dissipation of the servers and also increase the cooling requirements. Hence, workload-scheduling techniques have been proposed which intelligently place the workloads on available servers with the goal of saving power, reducing the temperature and the cooling requirements.

Nathuji and Schwan \cite{nathuji2007virtualpower} propose a dynamic power management technique to support the isolated and independent operation of VMs running on a virtualized platform, while globally coordinating the diverse power management strategies applied by the VMs to the virtualized resource. Their technique uses a set of virtualized power states to permit guest VMs to run their own, independent power management methods. Further, guest VM-level power management policies act upon these states to minimize power consumption, while meeting application requirements.

A thermal-aware technique for spatial workload placement in data centers is presented by Banerjee et al. \cite{banerjee2011integrating}. One of their techniques uses the information about dynamic behavior of computer room air conditioned (CRAC) to place the jobs in a manner which reduces the cooling demands from the CRACs. By integrating this technique with a temporal scheduling technique, the decision about when and on which server to execute a job can be taken, leading to a spatio-temporal scheduling technique.

Bradley et al. \cite{bradley2003workload} present a predictive power management technique for saving power in parallel computer systems. In web-based applications where the load on a system varies a lot, minimizing power consumption while meeting the demands of the workload is challenging. For such systems, their technique projects workload ahead of time to allow adequate resources to be powered on and kept ready for work when required to execute the workload. Their algorithm uses CPU utilization data to estimate the workload demand. When the utilization becomes larger than a predefined threshold, additional servers are powered on to reduce utilization on all servers to below that threshold. Conversely, when utilization at all servers is below that threshold and there is adequate capacity in the resulting server group to absorb the load of at least one server without any resource on any server being over-utilized, one or more server are powered off.

\subsection{Thermal-aware Power Management Techniques}

Several power management techniques work in thermal-unaware manner, i.e. they do not take the temperature-dependence of server power consumption into account. However, the heat-dissipation of processor components has strong temperature dependency and hence, an increase in operating temperature leads to increased heat-dissipation, which further increases the temperature and so on. This may drastically increase the component failure rates. To address this, several techniques have been proposed which manage the power consumption of the data centers while taking into account the thermal properties.

To minimize the total energy costs of data-center operation while providing a reliable thermal environment, Tang et al. \cite{tang2006thermal} propose a thermal-aware task-scheduling technique. In a blade server, multiple blades are integrated into each chassis, in which blades share a common supply and cooling fan. Each blade may itself have many processors. Operation of a chassis incurs chassis startup power consumption, along with actual power consumption of the blades. Hence, the power consumption cost of adding a task to one chassis may be different depending upon whether it involves waking up an idle chassis or an idle blade. Based on this observation, they evaluate several thermal-aware task-scheduling techniques, which exercise the trade-off between cost of start-up and power saving coming from turning off the blade servers.

Moore et al. \cite{moore2005making} present techniques to control the heat generation in servers using temperature-aware workload placement. Their techniques take into account the fact that since the heat may travel a large distance inside the data center before arriving at a temperature sensor, cooling inefficiencies can often arise in locations which are spatially uncorrelated from the place where the heat originates. Based on this, their technique uses the information about steady-state hot-spots and cold-spots in the data center to develop scheduling algorithms.

As discussed above, in data centers, work-scheduling algorithms have a significant influence on temperature distribution. Li et al. \cite{li2011thermocast} present a thermal forecasting model to predict the temperatures near the servers in a data center. Their model is based on continuous streams of temperature and airflow measurements. Their model integrates both physical laws and sensor observations in the data center. It uses the data obtained from the sensors to learn the parameters of the cyber-physical system of the data center.

\subsection{Other Techniques}

While most techniques aim to reduce energy consumption, a few techniques have also been proposed which aim to control peak power consumption. Lefurgy et al. \cite{lefurgy2007server} present a technique for controlling the peak power consumption of a high-density server. Their technique uses a feedback controller to periodically select the highest performance state while keeping the system within a fixed power constraint. The control loop is designed using the control theoretic methodology which helps in obtaining analytical guarantee on system stability and controller performance, despite variation in workloads. Their technique allows controlling power at different time-granularity, e.g. one second, eight seconds etc. The authors have shown that their technique performs better than the ad-hoc and open-loop techniques.

To fulfill the needs of several performance-critical and data-intensive applications that execute on many data center platforms \cite{gupta2008guaranteed,gupta2008mimo,mittal2011design,pande2009baywave}, a large amount of main memory resources need to be provisioned. Hence, improving the energy efficiency of main memories is important to achieving data center energy efficiency. Yoon et al. \cite{yoon2012boom} propose a technique for saving memory power consumption in data centers by intelligently utilizing low power mobile DRAM components. By using buffering mechanism to aggregate the data outputs from multiple ranks of low frequency mobile DRAM devices (such as 400MHz LPDDR2), their technique enables achieving high bandwidth and high storage capacity equal to server-class DRAM devices (such as 1600MHz DDR3). 

Chatterjee et al. \cite{chatterjee2012leveraging} propose a technique to take advantage of heterogeneity in DRAM memories to accelerate critical word access. In recent years, processors use DRAM chips with vastly differing latency and energy characteristics. Their technique recognizes the critical word in a cache line which is placed in a low-latency region of the main memory. The remaining non-critical words of the cache line are placed in a low-energy consuming region. Thus, their approach enables achieving high performance while reducing memory energy.

A technique to save the energy spent in unbeneficial refreshes in DRAM is proposed by J. Liu et al. \cite{liu2012raidr}. In a DRAM, only a small number of cells need to be refreshed at the minimum conservative refresh rate and the remaining cells can be refreshed at a much higher rate, while still maintaining their charge. Using this observation, their technique groups DRAM rows in multiple bins and uses different refresh interval for different bins. Further, by refreshing most of the cells with lower frequency than the leaky cells, they achieve reduction in the number of refresh operations required which leads to reduction in memory power consumption.

Keys et al. \cite{keys2012search} characterize a wide variety of clusters to find energy-efficient data center building blocks which are useful for data-intensive applications. The authors build homogeneous clusters using embedded, mobile and server systems and execute DryadLINQ applications on these clusters in order to study their energy efficiency for different application types. The authors have observed that their cluster was significantly more energy-efficient than ones with low-power server processors or embedded processors.

To maximize the benefits from the provided quality of computational services, while maintaining energy efficiency, Parolini et al. \cite{parolini2010cyber} present an energy saving technique for data centers. They model the data center in terms of mutually coupled cyber and physical systems. Here cyber component refers to the computational network representing the distribution and flow of computational tasks. The physical network refers to the thermal network which characterizes the distribution of thermal energy. The authors have shown that compared to the traditional approach in which the cyber and physical resources are controlled independently, their coordinated control strategy achieves more effective management of power consumption.

\section{Future Research Directions}

With CMOS scaling, the leakage energy consumption is increasing and hence the dynamic range of energy consumption that DVFS can utilize has reduced. Further, the design complexity of multicore processors also hinders straightforward use of DVFS for saving energy. For this reason, the effectiveness of DVFS based techniques has been reducing. Instead, the state-of-the-art in power management in data centers has focused on achieving energy-proportional computing by addressing hardware-platform heterogeneity \cite{delimitrou2013paragon}. Since modern data centers use tens to hundreds of servers with possibly different configurations, there exists significant heterogeneity in the computing servers. This heterogeneity is likely to increase further with increasing sizes of the data centers. Since workloads are sensitive to hardware platforms, a heterogeneity-oblivious scheduling of workloads on servers may lead to significant performance degradation. Thus, novel techniques have been proposed which take heterogeneity into account for scheduling workloads. Also, as discussed above, research efforts are also being directed towards utilizing renewable energy sources and dynamic power-state transitioning techniques. Also, researchers are exploring low-leakage high-density technologies such as non-volatile memory for reducing the energy consumption of computing systems \cite{mittal2013architectural}. These techniques reduce the carbon footprint of data centers, while also improving their energy efficiency.

It is expected that in the near future, several of the above mentioned solutions will be deployed in a synergistic manner for providing even larger savings. A key challenge for this is designing a coordination framework which is flexible and allows seamless integration of different techniques.

 \section{Conclusion}\label{sec:conclusion}
With the growing use of internet and requirement of data-storage and processing, the size of modern data centers has greatly increased. This has led to significant increase in the power consumption levels of the data centers. Moreover, the power consumption of data centers is approaching the limit imposed by thermal limitations of cooling solutions and power delivery. Also, since data centers are already consuming tens of Mega Watts, they are also stressing the capabilities of power generation facilities. 

As the complexity of operation of data centers increases, power management techniques which also ensure high-performance and low-costs are expected to become a crucial part of future enterprise architectures. In this paper, we highlighted the need for power management in data centers. We reviewed several techniques which have been proposed for reducing power consumption of data centers and classified them based on their characteristics. We believe that our survey will enable the researchers to gain insights into the state-of-the-art in power management of data centers and motivate them to propose innovative solutions for architecting future green data centers.

% Can use something like this to put references on a page
% by themselves when using endfloat and the captionsoff option.
\ifCLASSOPTIONcaptionsoff
  \newpage
\fi

% can use a bibliography generated by BibTeX as a .bbl file
% BibTeX documentation can be easily obtained at:
% http://www.ctan.org/tex-archive/biblio/bibtex/contrib/doc/
% The IEEEtran BibTeX style support page is at:
% http://www.michaelshell.org/tex/ieeetran/bibtex/
{
%\footnotesize
%\small
\bibliographystyle{IEEEtran}
\bibliography{PhDReferences}
}
% <OR> manually copy in the resultant .bbl file
% set second argument of \begin to the number of references
% (used to reserve space for the reference number labels box)

% that's all folks
\end{document}